\documentclass{emulateapj}

\usepackage{subfigure}
\usepackage[dvipsnames]{xcolor}
\usepackage{amsmath}

\newcommand{\myemail}{swjones@uvic.ca}

\newcommand{\iso}[2] {$^{#1}{\rm #2}$}          
\newcommand{\miso}[2] {^{#1}{\rm #2}}           
\newcommand{\plus}{& + &}                               
\newcommand{\makes}{& \rightarrow &}          
\newcommand{\msun}{$\,M_\odot\,$}               
\newcommand*\xbar[1]{                                      
  \hbox{%
    \vbox{%
      \hrule height 0.5pt 
      \kern0.5ex
      \hbox{%
        \kern-0.1em
        \ensuremath{#1}%
        \kern-0.1em
      }%
    }%
  }%
}

\slugcomment{Received 2014 March 31; accepted 2014 October 10; published 2014
December 2}

\shorttitle{Stars that ignite neon and oxygen off-center}
\shortauthors{Jones et al.}

\begin{document}

\title{The final fate of stars that ignite neon and oxygen off-center: electron
capture or iron core-collapse supernova?}

\author{Samuel Jones\altaffilmark{1,2,*}, Raphael Hirschi\altaffilmark{2,3},
and Ken'ichi Nomoto\altaffilmark{3,4}}
\affil{$^1$Department of Physics and Astronomy, University of Victoria, BC V8W 3P6, Canada\\
$^2$ Astrophysics Group, Lennard-Jones Building, Keele University ST5
5BG, UK \\
$^3$ Kavli Institute for the Physics and Mathematics of the Universe (WPI), The
University of Tokyo, Kashiwa, Chiba 277-8583, Japan \\
$^4$ Hamamatsu Professor
}

\altaffiltext{*}{email: \myemail}

\begin{abstract}
In the ONeMg cores of 8.8--9.5\msun stars, neon and oxygen burning is ignited
off-center. Whether the neon-oxygen flame propagates to the center is critical
to determine whether these stars undergo Fe core collapse or electron capture
induced ONeMg core collapse.  We present more details of stars that ignite neon
and oxygen burning off-center.  The neon flame is established in a similar
manner to the carbon flame of super-AGB stars, albeit with a narrower flame
width. The criteria for establishing a flame are able to be met if the strict
Schwarzschild criterion for convective instability is adopted.  Mixing across
the interface of the convective shell disrupts the conditions for the
propagation of the burning front and instead the shell burns as a series of
inward-moving flashes. While this may not directly affect whether the burning
will reach the center (as in super-AGB stars), the core is allowed to contract
between each shell flash. Reduction of the electron fraction in the shell
reduces the Chandrasekhar mass and the center reaches the threshold density for
the URCA process to activate and steer the remaining evolution of the core.
This highlights the importance of a more accurate treatment of mixing in the
stellar interior for yet another important question in stellar astrophysics -
determining the properties of stellar evolution and supernova progenitors at
the boundary between electron capture supernova and iron core-collapse
supernova.
\end{abstract}


\section{Introduction}
Differences in the pre-supernova structures of electron capture supernova
(EC-SN) and iron core-collapse supernova (FeCCSN) progenitors, and hence the
dynamics of the supernova explosion itself, are postulated to be responsible
for a number of observed phenomena. These phenomena include the orbital
eccentricity of BeX systems \citep{Knigge2011}, anti-correlations of silver and
palladium abundances with other elements which have relatively well-known
production sites \citep{Hansen2012}, type IIn-P supernovae
\citep{Anderson2012,Smith2013crab} and some faint branch supernovae with low
\iso{56}{Ni} ejecta.  In spite of this motivation the evolution of stars in the
mass range 8--10\msun, wherein lies the transition between stars that will end
their lives as EC-SNe and those that will produce FeCCSNe, is relatively
under-represented in the literature. So too under-represented is the evolution
of stars that ignite neon off-center
\citep[see][]{Woosley1980,Nomoto1980,Habets1986,NomotoHashimoto1988}. The
diversity of observations surrounding stars and supernovae as well as
improvements in the capabilities of numerical simulations have led to renewed
interest in these stars in recent years \citep{Umeda2012,Tauris2013}. These
stars have compact cores bounded by a steep density gradient along which the
supernova shock is accelerated. The boundary between EC-SN and FeCCSN is an
important ingredient for population synthesis and galactic chemical evolution,
however whether there are significant differences in the pre-supernova
structures of EC-SN and low-mass FeCCSN progenitors and hence whether there are
differences in their explosions and nucleosynthesis is still to be
investigated.

In our previous work \citep{Jones2013}, we presented new models of 8--12\msun
stars that suggest there could be an evolutionary channel producing electron
capture supernovae (EC-SNe hereafter) in addition to super-AGB stars. These
stars would be the most massive progenitors of EC-SNe and ignite neon and
oxygen burning off-center where the maximum temperature has moved outwards as
neutrino emission produces a net cooling of the mildly electron-degenerate
central regions. This off-center ignition of neon and oxygen is a
characteristic also exhibited by the lowest mass progenitors of iron
core-collapse supernovae (FeCCSNe).

It was suggested by \citet{Timmes1994} that should a star ignite neon
sufficiently far from its center, its fate could be an electron capture
supernova as opposed to an iron core-collapse supernova. The authors performed
simulations of nuclear flames in degenerate oxygen-neon (ONe) cores, providing
tabulated flame speeds as a function of temperature and density, however we
note that the lower bound of the table in density is still an order of
magnitude higher than the conditions under which neon is ignited in our
simulations.

In super-AGB stars, it is carbon that is ignited off-center. It has been shown
that in the limit of the strict Schwarzschild criterion for convection, a flame
is established and the nuclear burning front propagates inwards (towards the
center of the star). Under these assumptions, the flame migrates inwards from
the ignition point because the peak energy generation from \iso{12}{C} +
\iso{12}{C} fusion resides below the peak temperature
\citep{Siess2006,Denissenkov2013cflame}. This is due to the dependency of the
fusion rate on the square of the \iso{12}{C} abundance. Energy released by
carbon burning heats the material, dragging the peak temperature inwards and
thus the peak energy generation also moves inwards.

\begin{figure}
  \centering
  \includegraphics[width=1.\linewidth]{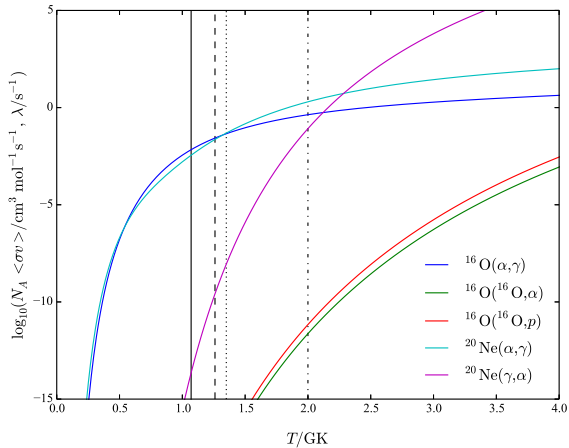}
  \caption[Rates of key neon- and oxygen-burning reactions as functions of
temperature from the REACLIB compilation]{Reaction rates, $\lambda$ (one-body
reactions) and $N_A<\sigma v>$ (two-body reactions) of key neon- and
oxygen-burning reactions as functions of temperature from the REACLIB
compilation \citep{Cyburt2010}. The vertical black lines are drawn in order to
facilitate comparison of the rates at temperatures of 1.07, 1.26, 1.35 and 2 GK
(see text for details).}
  \label{NeOflame_rates}
\end{figure}

\citet{Siess2009} studied the effect of thermohaline mixing on the evolution
and propagation of the carbon flame in super-AGB stars. Across the flame front,
there is a steep mean molecular weight gradient transitioning between the
unburnt composition (\iso{12}{C} and \iso{16}{O}) and the composition after it
has been processed by the flame (\iso{20}{Ne} and \iso{16}{O}). The
stratification of a fluid where material of higher mean molecular weight is
situated atop material of lower mean molecular weight can induce mixing,
depending on the steepness of the temperature gradient. Siess found that this
thermohaline mixing at the carbon flame front, if sufficiently strong, could
choke off the propagation of nuclear burning, and the carbon flame would thus
fail to reach the center of the star.

Stellar evolution calculations treating thermohaline mixing as a diffusive
process characterised by a salt-finger aspect ratio of $a\approx7$ are able to
reproduce the observed decrease of the surface \iso{12}{C} abundance and
\iso{12}{C}/\iso{13}{C} ratio in RGB stars \citep{Charbonnel2007}, however
recent two and three dimensional simulations of thermohaline mixing
\citep{Denissenkov2010,Traxler2011} have shown the mixing to be much less
efficient, characterised by a value of $a<1$. The simulations of
\citet{Siess2009} in which the carbon flame is quenched had also assumed the
same efficiency of thermohaline mixing, characterised by $a\approx7$.
\citet{Denissenkov2013cflame}, treating thermohaline mixing with the lower
efficiency determined from multi-dimensional hydrodynamics simulations, find
that this kind of mixing alone is not enough to quench the propagation of the
carbon flame in super-AGB stars, and the flame successfully reaches the center
of the star. Furthermore, Denissenkov and collaborators have tested the
resilience of the carbon-burning flame when assuming a convective boundary
mixing characterised by an exponentially decaying diffusion coefficient
\citep{Freytag1996,Herwig2000} with values of $f_{\rm CBM}=0.014$, 0.007 and
0.004 below the convective shell. In all cases, the flame was quenched, even
when accounting for the additional heat transport in the boundary mixing
region. The reason for the quenching of the flame in the presence of convective
boundary mixing is the flattening of the \iso{12}{C} abundance profile. The
conditions for the propagation of the flame are no longer satisfied and the
burning front does not reach the center. Under these assumptions, the super-AGB
star can produce a hybrid white dwarf, with an inner core of CO composition and
an outer core of ONe composition. \citet{Chen2014} have shown that the
quenching of the carbon flame in super-AGB stars leads to an increase in the
theoretical upper limit to the initial mass of type Ia supernova progenitors
when such hybrid white dwarfs are included. This could reduce the (already
narrow) initial mass range for which electron-capture supernovae are produced.

In this paper, we show the sensitivity of the neon-oxygen flame of low-mass
massive stars to mixing across the formal Schwarzschild boundary at the base of
the convection zone bounded by the flame. The mixing gives rise to a new
evolutionary path to electron capture supernovae in which the core
intermittently contracts between the recurrent quenching of neon-oxygen shell
burning. The star hence is able to reach central densities dominated by the
URCA-process, whose ability to remove electrons from the stellar plasma
accelerates the evolution of the star toward an electron capture supernova
(EC-SN), rather than an iron core collapse supernova (FeCCSN). For this reason,
we refer to these stars which ignite neon and oxygen off-center and produce an
EC-SN as failed massive stars (FMS).

\section{Input physics}
\label{methodsec}
The models shown in this paper have been computed using the stellar evolution
code MESA, with the same revision and input physics assumptions as in
\citet{Jones2013}, to which we refer the reader for more details. The only
exception is the treatment of weak interaction rates, for which we use the new
calculations of \citet{Toki2013} for the following three chains of weak
interactions:
\begin{equation}
\renewcommand{\arraystretch}{1.5}
\begin{array}{c l l l l l l l l l l l l l}
(i) & \miso{27}{Al}&\leftrightarrow&\miso{27}{Mg}&\leftrightarrow&\miso{27}{Na} \\ 
(ii) & \miso{25}{Mg}&\leftrightarrow&\miso{25}{Na}&\leftrightarrow&\miso{25}{Ne} \\ 
(iii) & \miso{23}{Na}&\leftrightarrow&\miso{23}{Ne}&\leftrightarrow&\miso{23}{F},
\end{array}
\end{equation}
which we showed in \citet{Jones2013} to make a non-negligible impact on the
evolution of stars achieving densities greater than $\rho\approx10^9\,{\rm
g\,cm}^{-3}$.  Traversing one link in the chain requires emission or absorption
of an electron or positron, and always releases a neutrino with a mean free
path much greater than the stellar radius. Calculating the electron fraction
accurately is, of course, very important when simulating the evolution of stars
that are supported by electron-degeneracy pressure. The strong neutrino cooling
arising when beta-equilibrium is established (URCA process) between a parent
and daughter nuclei from one such pair causes a steepening in the tail of the
electron distribution function. As a result, the impact of electron captures on
\iso{24}{Mg} and \iso{20}{Ne} that trigger the ignition of an oxygen
deflagration and the collapse of the core will be felt at higher densities.

\section{Ignition of neon and oxygen burning and development of a convective zone}
We previously presented 8.8\msun and 9.5\msun stellar models in which neon and
oxygen burning were ignited off-center in the degenerate core and proceeded to
burn inwards towards the stellar center \citep{Jones2013}. In the 8.8\msun
model, the burning failed to reach the center of the star and the calculation
resulted in an electron capture supernova. Conversely, the burning successfully
reached the center of the 9.5\msun model, which would become an iron
core-collapse supernova.

In our simulations of low-mass massive stars in which neon is ignited
off-center, the situation is a little more complicated than that of the carbon
flame in super-AGB stars. Rather than proceeding via the fusion of two similar
nuclei, neon-burning is driven by photodisintegration. The key reactions during
neon burning are
\begin{equation}
\renewcommand{\arraystretch}{1.5}
\begin{array}{c c c c c c c c c c c c c}
\miso{20}{Ne}\plus\gamma\makes\miso{16}{O}\plus\alpha \\
\miso{16}{O}\plus\alpha\makes\miso{20}{Ne}\plus\gamma\\
\miso{20}{Ne}\plus\alpha\makes\miso{24}{Mg}\plus\gamma.
\end{array}
\end{equation}
The neon photodisintegration reaction \iso{20}{Ne}($\gamma,\alpha$)\iso{16}{O}
has a Q-value of -4.73 MeV and is thus endothermic. When this reaction first
becomes significant, the inverse reaction,
\iso{16}{O}($\alpha,\gamma$)\iso{20}{Ne}, proceeds much faster, returning the
energy to the stellar material and replenishing the \iso{20}{Ne} abundance.
When the temperature becomes high enough however (see Fig.
\ref{NeOflame_rates}), the $\alpha$--particle released is quickly captured by
another \iso{20}{Ne} nucleus, producing \iso{24}{Mg}. This reaction has a
Q-value of 9.32 MeV and is the primary energy source during neon-burning. 

\begin{figure}
  \centering
  \includegraphics[width=1.\linewidth]{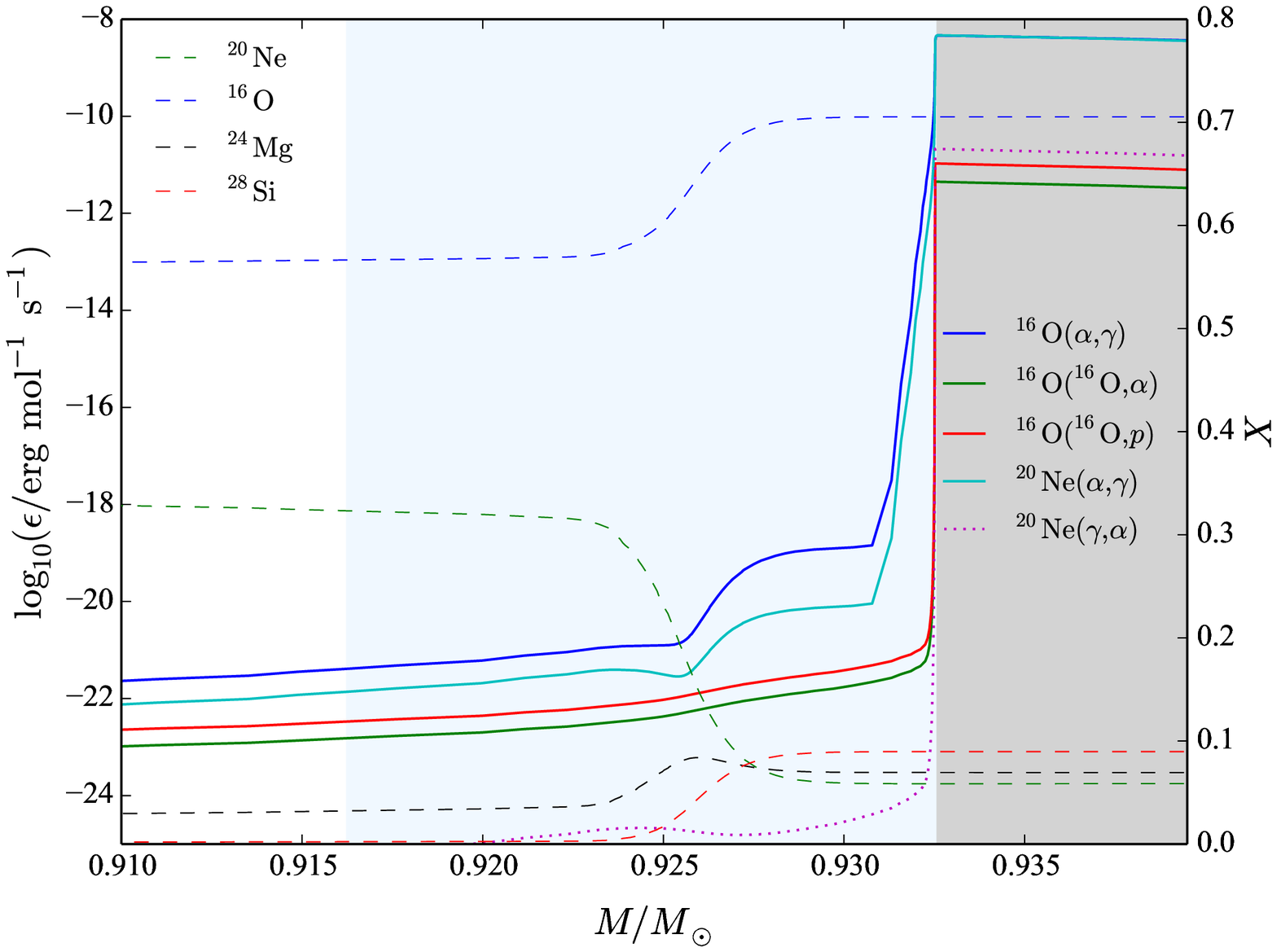}
  \includegraphics[width=1.\linewidth]{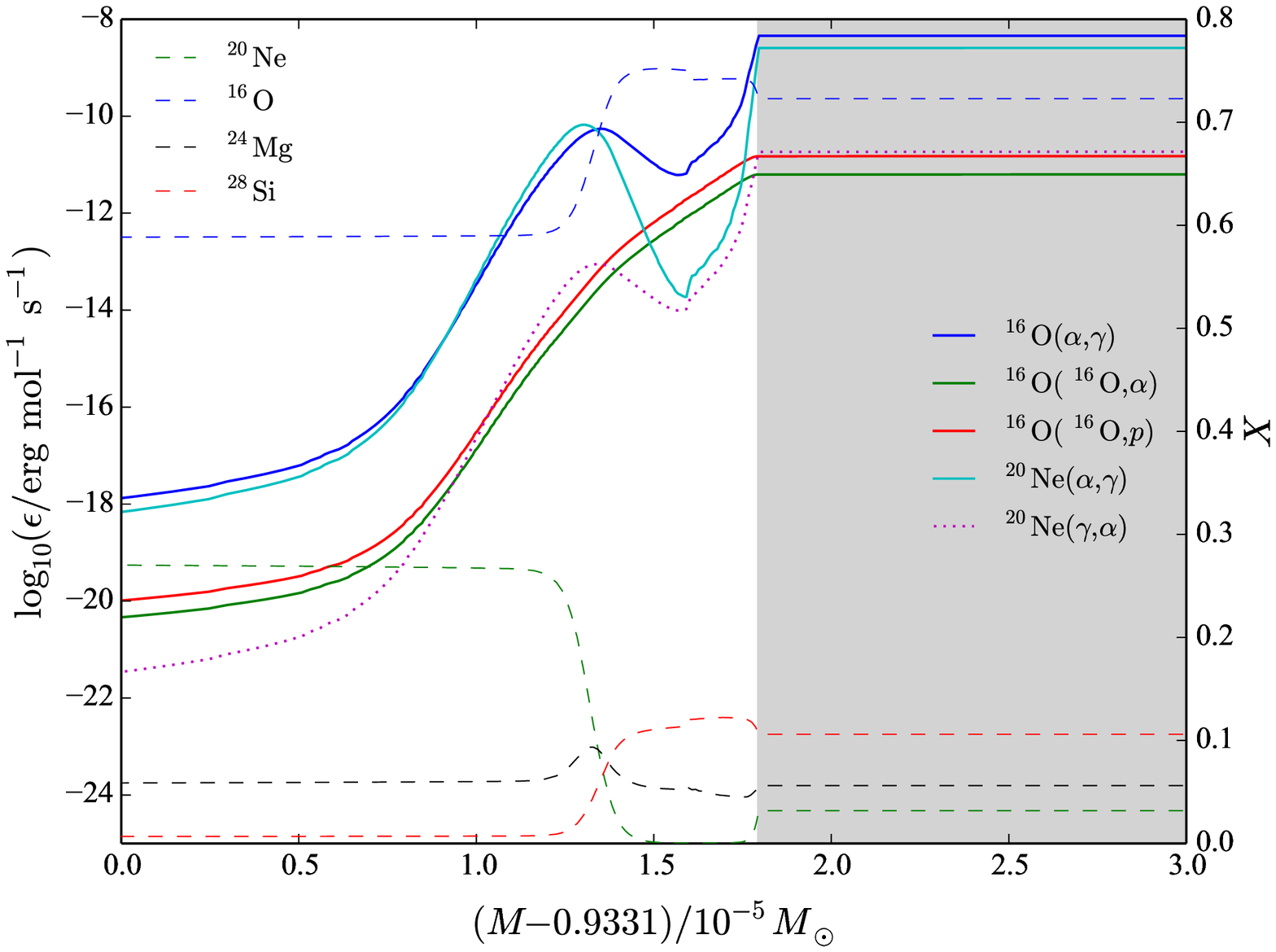}
  \caption[Energy production from key neon- and oxygen-burning reactions during
the peak of the first neon shell flash in the 8.8\msun model with different CBM
parameterisations]{Energy production from key neon- and oxygen-burning
reactions during the peak of the first neon shell flash in the 8.8\msun model
with $f_{\rm CBM}=0.005$ (top panel) and $f_{\rm CBM}=0$ (bottom panel) as
functions of mass coordinate (absolute values are plotted, with negative
quantities plotted with a dotted line style). The abundances of \iso{20}{Ne}
\iso{16}{O}, \iso{24}{Mg} and \iso{28}{Si} are plotted on the right axis. Note
the difference in the scale of the x-axis for the two plots. Regions of
convection are shaded grey and the extent of convective boundary mixing is
shaded for the $f_{\rm CBM}=0.005$ model (top panel) in light blue.}
  \label{Neflame_modeleps}
\end{figure}

\begin{figure*}
  \centering
  \includegraphics[width=.4\linewidth,clip=True,trim= 6mm 0mm 7mm 10mm]{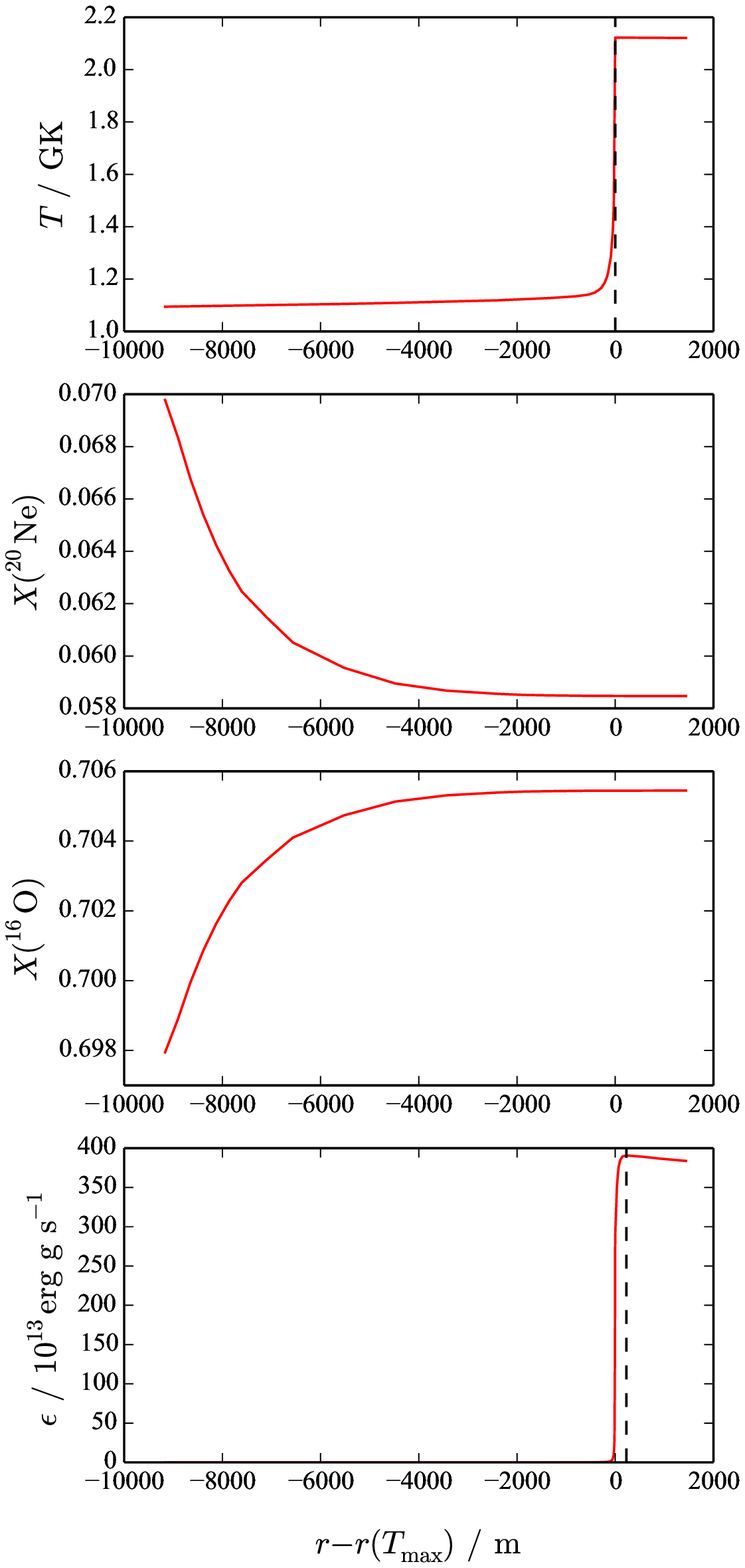}
  \includegraphics[width=.4\linewidth,clip=True,trim= 6mm 0mm 7mm 10mm]{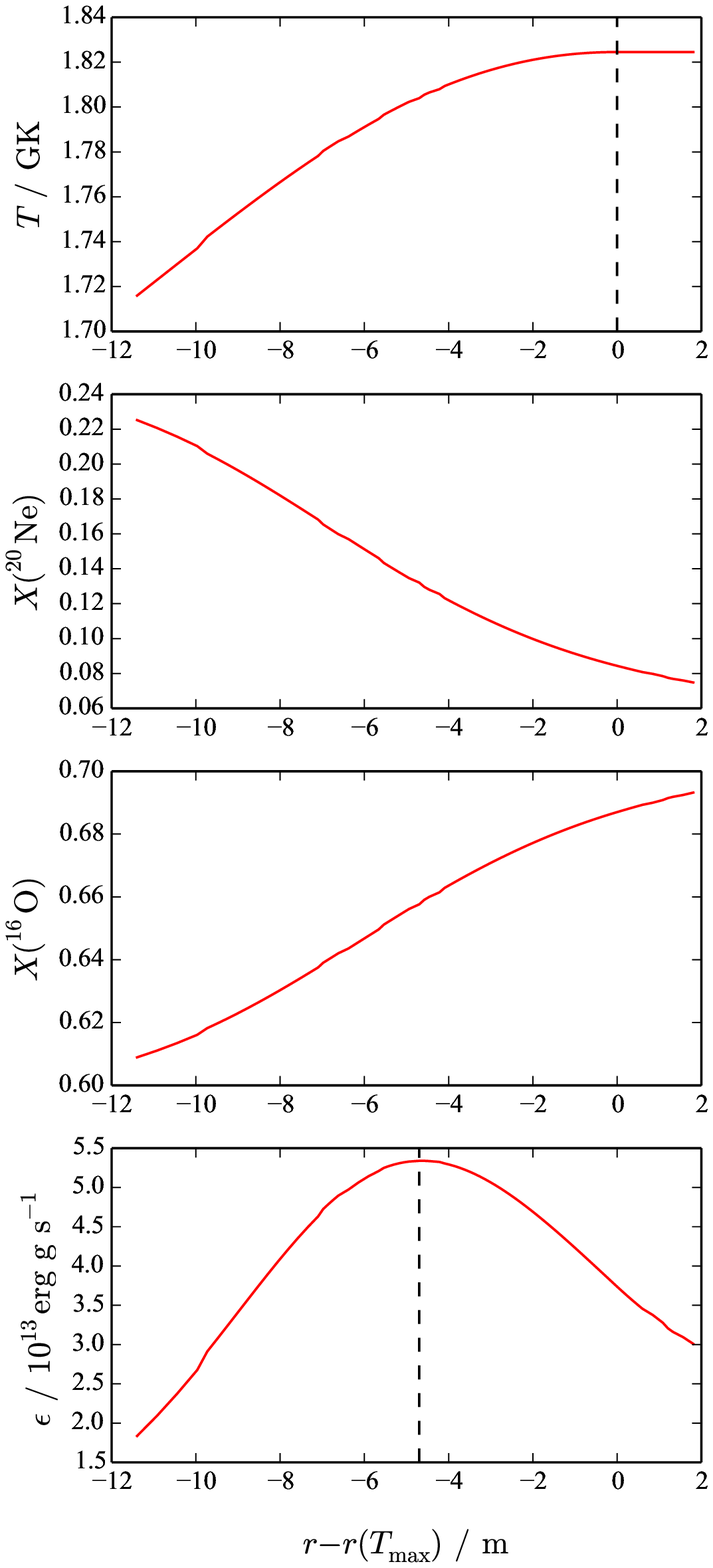}
  \caption{Properties of the model with ($f_{\rm CBM}=0.005$; left column) and
without ($f_{\rm CBM}=0$; right column) convective boundary mixing ($f_{\rm
CBM}=0$) in the region of peak nuclear energy production $\epsilon$ due to neon
burning. Dashed vertical lines show where quantities are a maximum. The peak
energy generation due to neon burning (bottom panels) clearly sits above the
peak temperature in the case with $f_{\rm CBM}=0.005$ (top left panel) and
below the peak temperature in the case with $f_{\rm CBM}=0$ (top right panel).
This difference is caused by the flattening of the \iso{20}{Ne} and \iso{16}{O}
abundance profiles by the convective boundary mixing (left center panels; c.f.
right center panels).}
  \label{quenchplot}
\end{figure*}

At the point where the heat accumulates, \iso{20}{Ne} is more efficient at
capturing the $\alpha$--particles released via the photodisintegration of neon.
The energy release is dictated by the photodisintegration rate and the burning
proceeds effectively as the net reaction
$2(\miso{20}{Ne})\rightarrow\,\miso{16}{O}+\,\miso{24}{Mg}+4.59$ MeV.  The
dependence of the energy generation rate on the neon abundance is not as
straightforward as for carbon burning, which is proportional to the square of
the carbon abundance. \citet{Arnett1974} and \citet{Woosley2002} have proposed
that the energy generation rate during neon burning scales like
$\epsilon\propto Y_{20}$ and $\epsilon\propto Y_{20}^2/Y_{16}$, respectively.
Both considerations omit the $\miso{24}{Mg}(\alpha,\gamma)\miso{28}{Si}$
reaction, which we find to be important and results in a scaling more like
$\epsilon\propto Y_{20}^{1.4}$ (see Appendix). In the Appendix, we employ a
steady-state alpha-particle  abundance approximation (which we qualify) in
order to look at the conditions under which the off-center neon-burning flame
is established in our stellar models and how the energy generation rate is
coupled to the stellar structure.

When the rate of energy generation from neon burning is high enough, the
material above the region of nuclear burning becomes convectively unstable.
When the fuel ignites off-center, like in the carbon flame of super-AGB stars
\citep{Siess2006}, the base of the convective zone does not develop at the
coordinate of the peak temperature, but a small distance above it. This is
because of the dependence of the luminosity on the temperature gradient,
$L_r\propto dT/dr$. Of course, $dT/dr=0$ at the coordinate of the peak
temperature and the criterion for convective instability,
$$L_r>\dfrac{16\pi acG}{3}\dfrac{mT^4}{\kappa P} \nabla_{\rm ad},$$
is satisfied at a point somewhere above the peak temperature where $\nabla_{\rm
rad}>\nabla_{\rm ad}$ (Schwarzschild criterion).

In the convective region a constant supply of fresh \iso{20}{Ne} is being mixed
down to the higher temperatures at the base. The temperature there is 1.26 GK,
where the \iso{16}{O}($\alpha,\gamma$)\iso{20}{Ne} and
\iso{20}{Ne}($\alpha,\gamma$)\iso{24}{Mg} reaction rates are incredibly similar
(see blue, cyan nd vertical black dashed lines in Fig. \ref{NeOflame_rates}).
As fresh \iso{20}{Ne} is mixed down to this region, the constant release of
energy raises the temperature enough to ignite oxygen burning, which proceeds
by the following reactions:
\begin{equation}
\renewcommand{\arraystretch}{1.5}
\begin{array}{c c c c c c c c c c c c c}
\miso{16}{O}\plus\miso{16}{O}\makes\miso{31}{P}\plus \miso{1}{H} \\
&&&&\miso{31}{P}\plus\miso{1}{H}\makes\miso{28}{Si}\plus\alpha\\
\miso{16}{O}\plus\miso{16}{O}\makes\miso{28}{Si}\plus\alpha \\
\end{array}
\end{equation}
The development of a convective zone provides luminosity to support the outer
layers of the core and temporarily halts the contraction of the core. The
central regions thus expand and cool. The lifetime of the shell burning episode
is lengthened while convection brings in fresh fuel to be burnt at the base of
the shell where the temperature is high.

\section{Propagation of the burning front towards the stellar center}
A large range of extra mixing extents at the convective boundary does not
affect the qualitative evolutionary outcome of the 8.8\msun model -- an EC-SN
\citep{Jones2013}. We now extend our analysis to include a case  in which there
is no convective boundary mixing across at the base of the neon-burning shell
(strict Schwarzschild criterion).

In the introduction, we summarised the recent works of \citet{Siess2009} and
\citet{Denissenkov2013cflame}, which show that mixing across the flame front
can destroy the conditions required for the persistence of a propagating
nuclear flame in the stellar interior (the radial stratification of the peak
energy production and temperature).  Fig. \ref{Neflame_modeleps} shows the
energy production due to the key neon- and oxygen-burning reactions during the
peak of the first neon shell flash episode in the 8.8\msun model. The top panel
is the case assuming $f_{\rm CBM}=0.005$ below the shell flash convection zone
and the bottom panel is for the case with $f_{\rm CBM}=0$ (no convective
boundary mixing). Note the difference in scale of the x-axis for the plots.

\begin{figure}
\centering
\includegraphics[width=1.\linewidth]{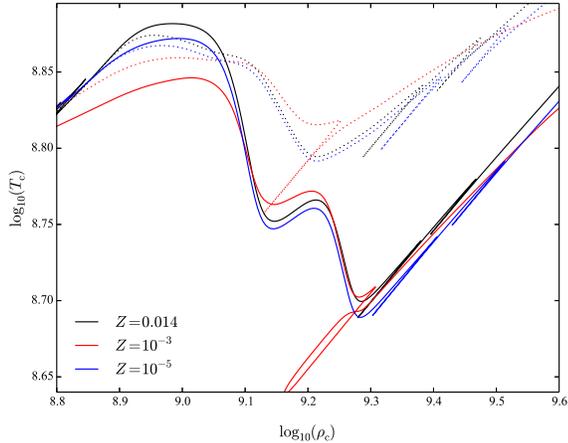}
\caption[Central evolution of representative failed massive star models at 3
metallicities ($Z=0.014$, 0.001 and $10^{-5}$) governed by the URCA
process]{Central evolution of representative failed massive star models at 3
metallicities ($Z=0.014$, 0.001 and $10^{-5}$) governed by the URCA process
during the neon shell-flash phase. Dotted lines show the models where the weak
rates of \citet{ODA94} were used and solid lines those where the new rates of
\citet{Toki2013} including Coulomb corrections were used.}
\label{allZ_toki_urca_tcrhoc}
\end{figure}

In the case with $f_{\rm CBM}=0$ (pure Schwarzschild criterion, bottom panel),
there are two distinct peaks in the energy production, separated by a thin
region strongly depleted in neon. Just below this region (to the left in the
plot), the temperature is about 1.35 GK and the
\iso{20}{Ne}($\alpha,\gamma$)\iso{24}{Mg} and
\iso{16}{O}($\alpha,\gamma$)\iso{20}{Ne} reaction rates are still very similar
(see blue, cyan and vertical black dotted lines in Fig.\,\ref{NeOflame_rates}).
The peak in energy production of each rate at this location traces the
abundance of the fuel, and so the peak in
\iso{20}{Ne}($\alpha,\gamma$)\iso{24}{Mg} lies just below that of
\iso{16}{O}($\alpha,\gamma$)\iso{20}{Ne}.

In the region where neon has been depleted, the temperature reaches 2 GK and
$\miso{16}{O} +\,\miso{16}{O} $ becomes significant. $\alpha$--particles are
released by \iso{16}{O}(\iso{16}{O}$,\alpha$)\iso{28}{Si} and
\iso{16}{O}(\iso{16}{O}$,p$)\iso{31}{P}($p,\alpha$)\iso{28}{Si}. As the
$\alpha$--particles are released in this way,
\iso{16}{O}($\alpha,\gamma$)\iso{20}{Ne} and
\iso{20}{Ne}($\alpha,\gamma$)\iso{24}{Mg} quickly turn \iso{16}{O} into
\iso{24}{Mg} and because the \iso{20}{Ne}($\alpha,\gamma$)\iso{24}{Mg} reaction
is much quicker than \iso{16}{O}($\alpha,\gamma$)\iso{20}{Ne} at this
temperature (vertical black dot-dashed line in Fig.\,\ref{NeOflame_rates}),
neon is completely depleted. \iso{24}{Mg}($\alpha,\gamma$)\iso{28}{Si} proceeds
at about half the rate of the \iso{20}{Ne}($\alpha,\gamma$)\iso{24}{Mg}
reaction and so the region starts to become enriched with \iso{24}{Mg} and
\iso{28}{Si}. It happens, then, that producing silicon from oxygen in this way
is quicker than oxygen-oxygen fusion, however it must rely upon the
oxygen-oxygen fusion reactions as the source of $\alpha$--particles.

After the neon is processed into \iso{24}{Mg}, \iso{28}{Si} and \iso{16}{O} by
the radiative pre-cursor neon flame, the burning moves inwards towards the
centre because of its strong dependence on the neon abundance.  Above the
neon-depleted region (to the right in the plot), neon-burning energy production
had previously boosted the luminosity above $L_{\rm crit}$ and the material is
convectively unstable, as we described earlier. The temperature in the
convective region increases and oxygen-burning reactions
($\miso{16}{O}\,+\,\miso{16}{O}$) are activated.

In the case with $f_{\rm CBM}=0.005$, the evolution up to the development of
the convective shell is the same for the case with $f_{\rm CBM}=0$, since there
is no mixing. However, the situation evolves differently once the shell becomes
convectively unstable. The extra mixing at the lower boundary of the convective
shell homogenises the composition across the thin radiative neon-burning shell
with that in the convective oxygen-burning shell. This mixing feeds the
convective shell with neon, which proceeds to burn there via the net reaction
$2(\miso{20}{Ne})\rightarrow\miso{16}{O}+\miso{24}{Mg}+4.59$ MeV at much higher
temperatures than are usually found during neon burning. This can be seen in
the higher rate of energy production by
\iso{20}{Ne}($\alpha,\gamma$)\iso{24}{Mg}---in particular, relative to that of
$\miso{16}{O}(\alpha,\gamma)\miso{20}{Ne}$---in the convective shell in the top
panel of Fig.\,\ref{Neflame_modeleps} ($f_{\rm CBM}=0.005$) compared to the
bottom panel ($f_{\rm CBM}=0$), and the higher abundances of \iso{20}{Ne} and
\iso{24}{Mg}.

In order for the flame to propagate, the energy production peak must be located
below the temperature peak. \citet{Denissenkov2013cflame} showed that in order
for the peak energy production to reside below the peak temperature during the
carbon-burning flame of super-AGB stars, the abundance gradient needs to be
sufficiently steep to overcome the large temperature exponent ($\epsilon_{\rm
nuc}\propto\rho X^2(\miso{12}{C})T^n$), where $n\approx40$.  In the absence of
convective-boundary mixing the neon-burning shell produces a steep gradient in
the abundance of neon just below the ignition point. As we will demonstrate in
Section~\ref{sec:neburn_appendix}, the abundance gradient of neon is steep
enough to establish a flame structure where the peak energy generation lies
below the peak temperature. The flame is much thinner than in the case of
off-center carbon ignition in a CO core and thus is more susceptible to
disruption by convective boundary mixing.

As discussed above (and is shown in greater detail in
Section~\ref{sec:neburn_appendix}), the peak in the energy generation of the
radiative pre-cursor neon flame follows closely the sharp increase in the
abundance of neon towards the centre (see the left column of
Fig.\,\ref{quenchplot} and compare to the right column of the same figure). If
there is mixing at the convective boundary between the radiative layer and the
convective shell, however, the step in the $X(\miso{20}{Ne})$ profile is (i)
smoothed out and (ii) displaced towards the centre of the star. With the mixing
assumed to be characterised by an exponentially decaying diffusion coefficient
with $f_{\rm CBM}=0.005$, the temperature at the new location of the step-up in
neon abundance towards the centre is 1.07 GK
($\log_{10}(T/\mathrm{K})\approx9.03$), and thus \iso{16}{O} dominates
\iso{20}{Ne} in the capturing of any $\alpha$--particles (see Fig.
\ref{NeOflame_rates}, vertical black solid line) and neon-burning barely
proceeds at all.

The shell burning episode continues to bring neon and oxygen into the
convective shell from the radiative layer below until both the convective
region and the region in which the mixing had extended are rich in
silicon-group composition (\iso{34}{S}, \iso{30}{Si} and \iso{28}{Si}, in order
of decreasing abundance by mass fraction) and depleted in \iso{16}{O} and
\iso{20}{Ne}. The convective shell persists until there is no longer sufficient
luminosity to sustain it. Upon the extinction of the convective shell, the core
contracts as described earlier. Neon burning re-ignites just below the extent
of the boundary mixing where fuel is abundant as the core heats up. This
behaviour, where the shell exhausts its fuel and extinguishes, is the critical
difference between the case with convective boundary mixing and the case
without.

\subsection{Sensitivity of the neon-burning flame to $f_{\rm CBM}$}
\label{sec:fcbm}
We have shown thus far that convective boundary mixing can disrupt the
abundance gradients and hence destroy the conditions under which a nuclear
flame can smoothly propagate through the degenerate oxygen-neon core.
Re-arranging the formula for the diffusion coefficient in the convective
boundary mixing region \citep{Freytag1996,Herwig1997} to give an expression for
$f_{\rm CBM}$ as
\begin{equation}
f_{\rm CBM} = -\dfrac{2z}{\lambda_{P,0}(\ln D - \ln D_0)},
\label{eq:Df_inv}
\end{equation}
we can estimate how small the $f_{\rm CBM}$ parameter can be before the flame
propagation may continue smoothly without intermittent quenching and
contraction. In Eq.~\ref{eq:Df_inv},~$\lambda_{P,0}$ is the scale height of
pressure at the lower boundary of the convective neon-burning shell and $D_0$
is the diffusion coefficient at that same location. In order to get an
order-of-magnitude estimate for the value of $f_{\rm CBM}$ that would disrupt
the neon-burning flame, we take the radial distance from the convective
boundary $z$ to be 10 m (approximately the width of the neon-burning flame
front in our simulations; see Fig.~\ref{quenchplot}). We also need a value for
the Diffusion coefficient $D$ at this location. In our models, the mixing
processes are frozen during neon-burning for diffusion coefficients less than
about $D=10^6$ cm$^2$ s$^{-1}$, and thus we adopt this value. In fact, the
value of $f_{\rm CBM}$ is not too sensitive to the exact value of $D$ and we
find that the value of $f_{\rm CBM}$ required to interrupt the flame
propagation is generally in the range $10^{-6}\lesssim f_{\rm CBM} \lesssim
10^{-5}$.

We have performed a similar calculation to the one above, this time for the
carbon flame in a super-AGB stellar model, the width of which is of the order
of 10 km. The formula in Eq.~\ref{eq:Df_inv} shows that in order for convective
boundary mixing (treated with the same diffusive model) to interfere with the
carbon flame, the minimum value of the convective boundary mixing parameter
would be $10^{-4}\lesssim f_{\rm CBM}\lesssim 10^{-3}$. This is consistent with
\citet{Denissenkov2013cflame}, who showed that the carbon flame is quenched
when $f_{\rm CBM}$ is as small as $7\times10^{-3}$.

\begin{figure}
\includegraphics[width=1.\linewidth]{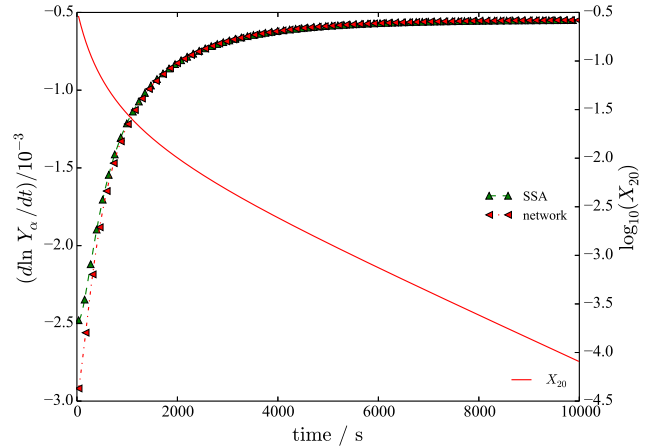}
\caption{Time-derivatives of the alpha-particle abundance as given directly by
a one-zone neon-burning network calculation (red sideways triangle) and by
assuming a steady-state abundance of alpha-particles for the same calculation
(green upright triangle). The formula for $d\ln Y_\alpha / dt$ under the
steady-state alpha-particle abundance approximation is given in
Eq.~\ref{eq:dlnya_SJ}.~$X_{20}$ is the mass fraction of \iso{20}{Ne} and is
drawn with a solid red line. The two time-derivatives agree well, showing that
a steady-state alpha-particle abundance is a good approximation when analyzing
the behaviour of neon burning.}
\label{fig:dYa_dt}
\end{figure}

\section{Electron capture supernova or iron core-collapse supernova?}
We have shown that mixing across the neon-burning flame front in stars that
ignite neon (and oxygen) off-center changes the manner in which neon burning
migrates toward the stellar center. With no mixing (assuming only the
Schwarzschild boundary), a flame is established and propagates (although we
have not simulated the conductive propagation all the way) to the center of the
star. When mixing across the convective boundary (more crucially, across the
flame front) takes place, the neon burning propagates in a discrete, step-wise
manner with intermittent periods of contraction. Because of the finite mass of
the fuel in the core, the number of shell flash episodes until the burning
reaches the center is a function of the depth of the convective boundary
mixing. This finding explains Fig.\,13 of \citet{Jones2013}, where the
simulations with larger values of $f_{\rm CBM}$ experience less flashes (sharp
deviations to the lower left of the figure) and stronger contraction following
each flash.

As long as the convective boundary mixing (for which we do not yet know the
strength) is strong enough to mix material across the flame front, a fraction
of stars that ignite neon and oxygen burning off-center will reach the
threshold density for the URCA process to be activated at their centers. These
stars could hence produce electron capture supernovae as their contraction is
accelerated toward the threshold density for electron captures by \iso{24}{Mg}
and \iso{20}{Ne} to be activated. In addition, the cooling caused by the URCA
process is stronger when using the new weak reaction rates calculated by
\citet[][see Fig.\,\ref{allZ_toki_urca_tcrhoc}]{Toki2013}. Each URCA pair
produces a cooling front that moves outwards from the center, leaving the
temperature a factor of 3 cooler than the threshold temperature for neon
ignition.

While the 9.5\msun model in our set \citep[see][]{Jones2013} assumed exactly
the same mixing assumptions as the 8.8\msun model, the neon and oxygen burning
had processed almost the entire core down to the center and as a result the
star would likely result in an FeCCSN. This is because the 9.5\msun model
ignites the off-center burning when its central density is much lower than in
the 8.8\msun model. As a result, the URCA process nuclei and the primary
electron-capture fuel \iso{20}{Ne} are destroyed before the threshold density
for the URCA process is reached at the center. The slope of the evolution in
the $\rho_{\rm c}-T_{\rm c}$ plane during the shell neon-burning phase (shown
in Fig.\,\ref{allZ_toki_urca_tcrhoc}) is close to the adiabatic one, meaning
that the same slope is found in all of the models. This means that if
activation of the URCA process in these stars will indeed always produce an
electron capture supernova, then there could be a relationship between the
central density of the star at the time of off-center neon ignition and the
fate of the star -- EC-SN or FeCCSN.

\begin{figure}
\includegraphics[width=1.\linewidth]{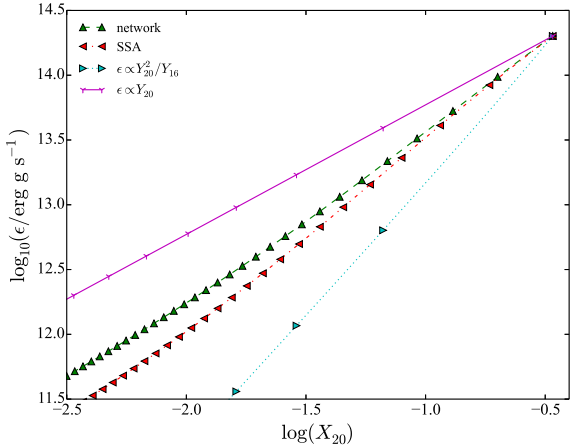}
\caption{Energy generation rate during neon burning as a function of the
\iso{20}{Ne} mass fraction. The numerically-evaluated rate from the network
calculation is shown in green (upright triangles) along with the energy
generation predicted by the analytical formula in Eq.~\ref{eq:newepseq}
assuming a steady-state abundance of alpha-particles (red sideways triangles).
For comparison, we show the energy generation rates given by the relations in
\citet{Woosley2002} and \citet{Arnett1974} which scale as $\epsilon\propto
Y_{20}^2/Y_{16}$ and $\epsilon\propto Y_{20}$, respectively.}
\label{fig:eps_many_methods}
\end{figure}

\section{Discussion and concluding remarks}
\label{conclusions}
In this paper, we have shown more details of stars that ignite neon and oxygen
burning off-center in a degenerate core. If hydrodynamical instabilities at the
base of the convection zone trailing the narrow flame front induce mixing
across the front then the nuclear burning migrates toward the center in a
step-wise manner with intermittent periods of contraction. If there is strictly
no mixing across the front, then the propagation ensues as a flame chasing the
abundance gradient of \iso{20}{Ne}.  For a convective boundary mixing model
based on an exponentially decaying diffusion coefficient
\citep[e.g.][]{Freytag1996,Herwig1997}, quenching of the flame can be achieved
for diffusion coefficient e-folding lengths of around $10^{-6}\lambda_{P}$ or
larger.  The equivalent e-folding length for quenching of the carbon flame in
super-AGB stars is greater by about three orders of magnitude.  This e-folding
length of the diffusion coefficient required to quench the neon flame could be
considered small in comparison and thus (if the neon flame is subjected to the
same types of hydrodynamical instabilities as the carbon flame) if the carbon
flame can be quenched, it is very likely that neon flame will experience
quenching.  Of course, these results are limited by the assumption that the
convective boundary mixing model that we have adopted is representative of the
hydrodynamic processes that induce the mixing at such a boundary in stars.  3D
hydrodynamics simulations of the neon-flame, capturing the turbulent motion
such as the ones presented in \citet{Meakin2007}, \citet{Herwig2011,Herwig2014}
and \citet{Viallet2013}, are needed to better constrain the width of the flame
and more importantly the type of mixing that occurs across the boundary.

We have shown that it is the intermittent periods of contraction due to the
quenching of the burning front by convective boundary mixing that cause an
8.8\msun model to reach the threshold density for the URCA process to be
activated. The model contracts to higher densities following each shell flash
owing to the reduction in the electron fraction in the oxygen-burning shell.
The ultimate removal of electrons by the URCA process drive the evolution
towards an electron capture supernova.

In order to progress further with this study and make predictions about how
failed massive stars may contribute to the EC-SN rate, we require constraints
on the behaviour of both the flame itself and the strength of mixing across the
flame front. Multi-dimensional hydrodynamics simulations should give the
insight required to make a firm conclusion. If indeed, the mixing is strong
enough to cause the step-wise propagation of the flame, the results of
\citet{Tauris2013} suggest that electron capture supernovae from stars in
binary configurations could even produce type Ic explosions.

\appendix
\section{Discussion and derivations of neon-burning equations}
\label{sec:neburn_appendix}
The rate of change of the helium abundance during neon burning is given by the expression
\begin{equation}
\dot{Y}_\alpha = -Y_\alpha(Y_{20}\lambda_{20\alpha\gamma} + Y_{16}\lambda_{16\alpha\gamma} + Y_{24}\lambda_{24\alpha\gamma}) + Y_{20}\lambda_{20\gamma\alpha} + Y_{24}\lambda_{24\gamma\alpha}.
\label{eq:Yadot}
\end{equation}
If the alpha abundance is in a steady state, then $\dot{Y}_\alpha\approx0$ and
Equation~\ref{eq:Yadot} then gives an expression for the alpha-particle
abundance as
\begin{equation}
Y_\alpha=\dfrac{Y_{20}\lambda_{20\gamma\alpha}+Y_{24}\lambda_{24\gamma\alpha}}{Y_{20}\lambda_{20\alpha\gamma} + Y_{16}\lambda_{16\alpha\gamma} + Y_{24}\lambda_{24\alpha\gamma}}.
\label{eq:SSalpha_abundance}
\end{equation}
The quantities $\lambda$ are the reaction rates with units of either g~s$^{-1}$
($\rho N_{\rm A} <\sigma v>$) for the two-body reactions or just s$^{-1}$ for
the photo-disintegration reactions. At neon-burning temperatures of about
1.8~GK, $\lambda_{20\gamma\alpha}/\lambda_{24\gamma\alpha}\approx10^{10}$ and
thus the $\miso{24}{Mg}(\gamma,\alpha)\miso{20}{Ne}$ reaction can be considered
negligible.
Taking the logarithm and then the time derivative of both sides of
Equation~\ref{eq:SSalpha_abundance} yields the relation
\begin{equation}
\dfrac{d\ln Y_\alpha}{dt} = \dfrac{d\ln Y_{20}}{dt}-\dfrac{d}{dt}\ln (Y_{20}\lambda_{20\alpha\gamma} + Y_{16}\lambda_{16\alpha\gamma} + Y_{24}\lambda_{24\alpha\gamma}).
\label{eq:dlnya_SJ}
\end{equation}
In Fig.~\ref{fig:dYa_dt} $d\ln Y_\alpha/dt$ calculated using the formula in
Eq.\,\ref{eq:dlnya_SJ} is shown for a one-zone nucleosynthesis simulation at
$T_9=1.82$ GK and $\rho=2.79\times10^7{\rm g~cm}^{-3}$, along with the actual
value of $d\ln Y_\alpha/dt$ computed by the network. These are the conditions
under which the neon shell burns in our 8.8\msun stellar model.
Fig.~\ref{fig:dYa_dt} shows that the steady-state alpha-particle abundance
approximation is indeed appropriate for the analysis of neon-burning behaviour.

It is important to note at this point that a steady-state abundance of
alpha-particles is not assumed within the reaction network or any other part of
the MESA code, which performs a network integration that is fully coupled to
the structure and mixing operators. We use the steady-state alpha-particle
abundance only as a means to describe the behaviour of neon burning---and in
particular its interaction with convective boundary mixing---in the stellar
models that we have calculated.

Following the derivation of \cite{Arnett1974}, under the assumption of a
steady-state alpha-particle abundance, the time derivatives of the other
abundances are given by
\begin{equation}
\renewcommand{\arraystretch}{1.5}
\begin{array}{c c l c c c c c }
\dot{Y}_{16} & = & Y_{20}\lambda_{20\gamma\alpha}(1-A), \\
\dot{Y}_{20} & = & Y_{20}\lambda_{20\gamma\alpha}(A - 1 - B), \\
\dot{Y}_{24} & = & Y_{20}\lambda_{20\gamma\alpha}(B - C), \\
\dot{Y}_{28} & = & Y_{20}\lambda_{20\gamma\alpha}C,
\end{array}
\label{eq:abund_time_derivs}
\end{equation}
where $A$, $B$ and $C$ are defined as
\begin{equation}
\renewcommand{\arraystretch}{2.5}
\begin{array}{c c l c c c c c }
A & = & \dfrac{Y_{16}\lambda_{16\alpha\gamma}}{\xi} \\
B & = & \dfrac{Y_{20}\lambda_{20\alpha\gamma}}{\xi} \\
C & = & \dfrac{Y_{24}\lambda_{24\alpha\gamma}}{\xi}.
\end{array}
\label{eq:abcdef}
\end{equation}
and $\xi$ is the denominator in the RHS of Eq.~\ref{eq:SSalpha_abundance},
\begin{equation}
\xi = Y_{20}\lambda_{20\alpha\gamma} + Y_{16}\lambda_{16\alpha\gamma} + Y_{24}\lambda_{24\alpha\gamma}.
\end{equation}
An expression for the energy generation during neon-burning
($\epsilon\propto\sum_i \dot{Y}_iB_i$, where $B_i$ is the binding energy of a
nucleus of type $i$) can now be formulated using the time-derivatives of the
key abundances (Eq.~\ref{eq:abund_time_derivs}) and the fact that $A+B+C=1$
(which we have also confirmed numerically). The resulting expression is
\begin{equation}
\ln \epsilon=\ln Y_{20} + \ln \lambda_{20\gamma\alpha} + \ln\phi - \ln\zeta + D
\label{eq:newepseq}
\end{equation}
with
\begin{equation}
\phi = NY_{24}<\sigma v>_{24\alpha\gamma} + [N+M]Y_{20}<\sigma v>_{20\alpha\gamma},
\label{eq:phidef}
\end{equation}
$D$ is a constant and $\zeta=\xi/\rho$. In Fig.~\ref{fig:eps_many_methods} we
show the energy generation rate given by Eq.~\ref{eq:newepseq} compared to the
rate evaluated numerically by the reaction network. We also show in
Fig.~\ref{fig:eps_many_methods} the scaling of the neon-burning energy
generation with the \iso{20}{Ne} abundance as formulated by \citet{Woosley2002}
and \citet{Arnett1974}, which scale as $\epsilon\propto Y_{20}^2/Y_{16}$ and
$\epsilon\propto Y_{20}$, respectively and fail to reproduce the trend in our
network calculation. The peak energy generation for each approximation has been
normalised to the peak energy generation from the reaction network in order to
more clearly highlight the differences between the different cases. Although we
follow an almost identical derivation to \citet{Arnett1974}, the main
difference is that Arnett assumes that the
$\miso{24}{Mg}(\alpha,\gamma)\miso{28}{Si}$ reaction is negligible (i.e. $C=0$
in Eq.~\ref{eq:abcdef}), which is not the case in our calculations. From the
network calculation, the slope down to about $X_{20}=0.01$ is close to
$\epsilon\propto Y_{20}^{1.4}$.

In order for the conditions for a flame structure to be established, it must be
possible that
\begin{equation}
\dfrac{\partial\ln\epsilon}{\partial\ln T} = \dfrac{\partial\ln Y_{20}}{\partial\ln T} + \dfrac{\partial\ln\lambda_{20\gamma\alpha}}{\partial\ln T} + \dfrac{\partial\ln\phi}{\partial\ln T} - \dfrac{\partial\ln\zeta}{\partial\ln T} < 0.
\label{eq:dlneps_dlnT}
\end{equation}
We have evaluated each derivative in Eq.~\ref{eq:dlneps_dlnT} numerically from
our stellar model and although none of the terms are negligible, it is the
first two terms (involving $Y_{20}$ and $\lambda_{20\gamma\alpha}$) that
dominate. $\partial\ln\lambda_{20\gamma\alpha}/\partial\ln T$ remains
reasonably fixed around the flame location at about 42. The abundance gradient
of \iso{20}{Ne} resulting from the thin flame front is steep enough to dominate
over the gradient of the $\miso{20}{Ne}(\gamma,\alpha)\miso{16}{O}$ rate and
together with the other terms results in $\partial\ln\epsilon/\partial\ln T<0$
across a region of a few meters.  The flame width is much thinner for
off-center neon burning compared to off-center carbon burning, making it more
susceptible to destruction by convective boundary mixing (see
Section~\ref{sec:fcbm}).

\acknowledgments
The research leading to these results has received funding from the European
Research Council under the European Union's Seventh Framework Programme
(FP/2007-2013)/ERC Grant Agreement n. 306901.  R. H. thanks the Eurocore
project Eurogenesis for support.  K. N., R. H. and S. J. acknowledge support
from the World Premier International Research Center Initiative (WPI
Initiative), MEXT, Japan.

\bibliographystyle{apj}
\bibliography{references}

\end{document}